\documentclass[a4paper]{article}

\usepackage{amsthm,amsmath,amssymb}
\usepackage[utf8]{inputenc} 
\usepackage{graphicx}
\usepackage{subfigure}
\usepackage{pgfplots,pgfplotstable}
\usepackage{tikz,environ}
\usepackage[normalem]{ulem}
\usetikzlibrary{plotmarks}
\usepackage{bm}
\usepackage{geometry} 
\newtheorem{hypo}{Assumption}[section]

\newtheorem{prop}[hypo]{Proposition}

\newcommand{\eps}{\varepsilon}

\newcommand{\fX}{{\mathfrak X}}

\newcommand{\vfv}{\bm{\mathfrak v}}

\newcommand{\fp}{{\mathfrak p}}

\newcommand{\fr}{{\mathfrak r}}

\newcommand{\fy}{{\mathfrak y}}
\newcommand{\fz}{{\mathfrak z}}

\newcommand{\IR}{{\mathbb R}}
\newcommand{\IC}{{\mathbb C}}

\newcommand{\ra}{{\mathrm a}}
\newcommand{\rb}{{\mathrm b}}
\newcommand{\rd}{{\mathrm d}}
\newcommand{\rh}{{\mathrm h}}

\newcommand{\Rd}{R_{\textrm{d}}}

\newcommand{\bDelta}{\boldsymbol{\Delta}}
\newcommand{\ve}{{\mathbf e}}
\newcommand{\vf}{{\mathbf f}}

\newcommand{\vn}{{\mathbf n}}

\newcommand{\vx}{{\mathbf x}}

\newcommand{\vv}{{\mathbf v}}

\newcommand{\vV}{{\mathbf V}}
\newcommand{\vX}{{\mathbf X}}

\newcommand{\nB}{{\mathcal B}}
\newcommand{\nD}{{\mathcal D}}
\newcommand{\nN}{{\mathcal N}}

\newcommand{\fnV}{\bm{\mathcal V}}
\newcommand{\nP}{{\mathcal P}}
\newcommand{\zerobf}{\boldsymbol{0}}

\newcommand{\Ltwo}       {\mathrm{L}^2}

\newcommand{\Hone}       {\mathrm{H}^1}
\newcommand{\Honeloc}    {\mathrm{H}_{\text{loc}}^1}

\newcommand{\inc}       {\mathsf{inc}}
\newcommand\xnorm[2]{\left\lVert #1 \right\rVert_{#2}}

\newcommand{\imag}{\ensuremath{{\rm i}}}
\DeclareMathOperator{\Laplace}{\Delta}
\DeclareMathOperator{\Div}{div}
\newcommand{\Curl}{\operatorname{{\bf curl}}}
\DeclareMathOperator{\realpart}{Re}
\renewcommand{\Re}{{\realpart}}
\DeclareMathOperator{\imagpart}{Im}
\renewcommand{\Im}{{\imagpart}}

\newcommand\liner{\text{liner}}
\newcommand\hole {\text{hole}}

\newcommand\ie{\emph{i.e.},\,}
\newcommand\eg{\emph{e.g.},\,}

\newcommand\bjump[1]{\big[ #1 \big]}
\newcommand\bavrg[1]{\big< #1 \big>}

\begin{document}

\begin{center}

{\Huge{\sffamily\bfseries On impedance conditions for circular\\[0.1em] multiperforated acoustic liners}}\\[0.5cm]

{\large{\sffamily Kersten Schmidt$^a$, Adrien Semin$^{b}$, Anastasia
    Thöns-Zueva$^c$, Friedrich Bake$^d$}}\\[0.5cm]

{\sffamily\small $a$: Technische Universität Darmstadt, Fachbereich
  Mathematik, AG Numerik und Wissenschaftliches Rechnen, Dolivostrasse
  15, 64293 Darmstadt,Germany}\\ 
{\sffamily\small $b$: Branderburgische Technische Universität
  Cottbus-Senftenberg, Institut für Mathematik, Platz der Deutschen
  Einheit 1, 03046 Cottbus, Germany}\\
{\sffamily\small $c$: Institut f\"ur Mathematik, Technische Universit\"at
  Berlin, Stra\ss e des 17. Juni 136, 10623 Berlin, Germany} \\
{\sffamily\small $d$: German Aerospace Center, Institute of Propulsion
  Technology, Müller-Breslau-Stra\ss e 8, 10623 Berlin, Germany} \\
\end{center}

\noindent \textbf{\Large\sffamily Abstract}\\

\vspace{-2em}
\paragraph{\sffamily Background} 

The acoustic damping in gas turbines and aero-engines relies to a
great extent on acoustic liners that consists of a cavity and a
perforated face sheet. %
The prediction of the impedance of the liners by direct numerical
simulation is nowadays not feasible due to the hundreds to thousands
repetitions of tiny holes.  We introduce a procedure to numerically
obtain the Rayleigh conductivity for acoustic liners for viscous gases
at rest, and with it define the acoustic impedance of the perforated
sheet.

\vspace{-1em}
\paragraph{\sffamily Results} 

The proposed method decouples the effects that are dominant on
different scales: (a) viscous and incompressible flow at the scale of
one hole, (b) inviscid and incompressible flow at the scale of the
hole pattern, and (c) inviscid and compressible flow at the scale of
the wave-length.  With the method of matched asymptotic expansions we
couple the different scales and eventually obtain effective impedance
conditions on the macroscopic scale.  For this the effective Rayleigh
conductivity results by numerical solution of an instationary Stokes
problem in frequency domain around one hole with prescribed pressure
at infinite distance to the aperture.  It depends on hole shape,
frequency, mean density and viscosity divided by the area of the
periodicity cell.  This enables us to estimate dissipation losses and
transmission properties, that we compare with acoustic measurements in
a duct acoustic test rig with a circular cross-section by DLR Berlin.

\vspace{-1em}
\paragraph{\sffamily Conclusions} 
  
A precise and reasonable definition of an effective Rayleigh
conductivity at the scale of one hole is proposed and impedance
conditions for the macroscopic pressure or velocity are derived in a
systematic procedure.  The comparison with experiments show that the
derived impedance conditions give a good prediction of the dissipation
losses.

\noindent \paragraph{\sffamily Keywords}
Acoustic liner, Perforated plates, Multiscale analysis, Rayleigh
conductivity, Impedance conditions.

\noindent \paragraph{\sffamily MSC classification codes}
35Q30, 35B27, 74Q15, 76M50




\section{\sffamily Introduction}

The safe and stable operation of modern low-emission gas turbines and
aero-engines crucially depends on the acoustic damping capability of the
combustion system components. Hereby, so called bias flow liner -- consisting
of a cavity and a perforated face sheet with additional cooling air flow --
play a significant role. Since decades the damping performance prediction of
these bias flow liner under all possible flow conditions remains a major
challenge. However, due to the higher tendency of low-emission, lean burn
combustion concepts for combustion instabilities the prediction of the acoustic
bias flow liner impedance and therewith its damping performance is a very
important prerequisite for the engine design process.  Several analytical and
semi-empirical models for the impedance description of bias flow liner were
developed in the past (see also~\cite{Lahiri:2014}). This work focuses on the
numerical simulation of the acoustic characteristics of bias flow liner
applying multi-scale modeling.

In principal all theoretical approaches are based on the formulation of the Rayleigh conductivity $K_R$~\cite{rayleigh1871,rayleigh1945} 
which describes the ratio of the fluctuating volume flow $Q(t)$ through a hole to the driving pressure difference $P^-(t)-P^+(t)$ across the hole:
\begin{equation}
  \label{eq:KR_literature}
  K_R := \frac{\rho_0 \partial_t Q(t)}{P^-(t)-P^+(t)},
\end{equation}
and has the dimensions of length.  One major challenge in the model
description of the Rayleigh conductivity represents the definition or
the specification of the pressure difference since, above and below the
perforated liner face-sheet the pressure is not necessarily constant
rather a function of the distance from the hole.  Here, the present
work applying a multiscale asymptotic model will provide an exactly
defined solution.  More precisely, the Rayleigh conductivity of a
single hole in an array of holes is distributed over the whole liner
area.  In this way the {\em effective Rayleigh conductivity}
\begin{align}
  k_R = \frac{K_R}{A_\delta}
\end{align}
as quotient of the Rayleigh conductivity of one hole and the area
$A_\delta$ of one periodicity cell of the array is introduced that has
the dimensions of one over length.  Using the effective Rayleigh
conductivity the liner impedance can be determined like later shown
for example in equation~\eqref{eq:impedance_our_model}.

\section{\sffamily Methods}

We consider acoustic liner that consist of a wall or part of a wall
with a periodic dense array of equisized and equishaped holes with an
characteristic periodicity that is proportional to a small parameter
$\delta > 0$.  The holes may not be of cylindrical shape and even
tilted in general. %
For sake of simplicity we consider the perforated wall
$\Omega_\liner^\delta$ with a circular cross-section of inner radius
$\Rd$, %
while noting that the proposed procedure to define the Rayleigh
conductivity and impedance conditions do not depend on the choice of
the cross-section, but only on the hole pattern and hole shape and can
be directly transfered to other cross-sections like rectangular. %
 
\begin{figure}[!bt]
  \centering
  \includegraphics[width=0.95\linewidth]{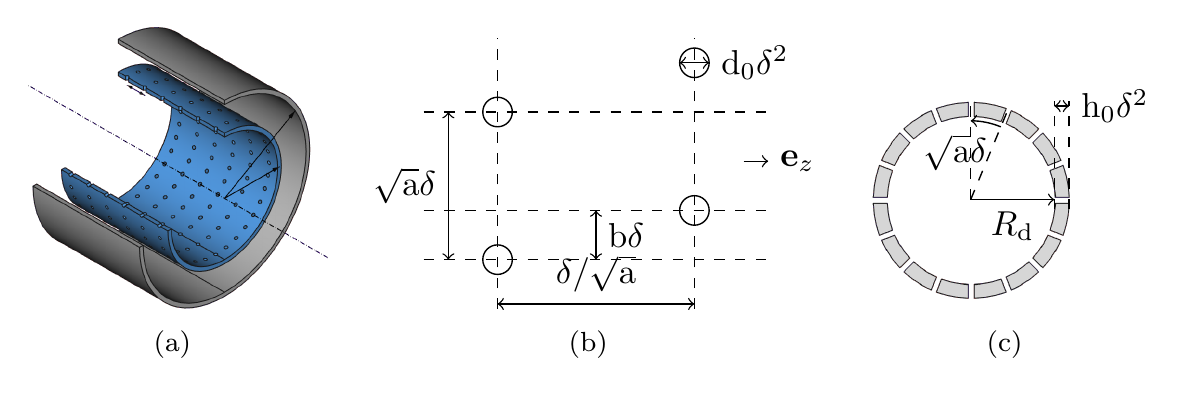}
  \caption{(a) Simplified geometry of a combustion liner for
    acoustic studies. (b) flattened liner. (c) view along a cross-section.}
  \label{fig:liner_geometry}
\end{figure}

To derive the impedance conditions we let the parameter $\delta$ of
the hole period tending zero -- so the number holes increases
accordingly -- while the inner and outer diameter of the cross section
are scaled like $\delta^2$ as well as the thickness of the perforated
wall, see Fig.~\ref{fig:liner_geometry}.  As $\delta \to 0$, the holes
merge and the domain $\Omega^\delta_\liner$ degenerates to an
interface $\Gamma_\liner$ on which we will prescibe the impedance
conditions representing the correct disspation behaviour of the
acoustic liner. For the circular liner the limit interface domain
$\Gamma_\liner$ is an cylinder of radius~$\Rd$. %
As it simplifies the derivation and impedance condition greatly we
assume that the area of the periodicity cell of the periodic array
$A_\delta = \delta^2$.

This liner shall be embedded in a duct domain $\Omega$ and the
computational domain is
$\Omega^\delta := \Omega \setminus \Omega_\liner^\delta$ for every
$\delta > 0$, \ie, the duct domain without the multi-perforated wall.
On this domain we introduce as viscoacoustic model %
the linearized compressible Navier-Stokes equations in frequency
domain in a uniform and stagnant media for a source term %
$\vf(t,x) = \Re(\vf(x) \exp(-\imath \omega t))$ with an angular
frequency $\omega > 0$:
\begin{subequations}
  \label{eq:Navier_Stokes}
  \begin{align}
    - \imath \omega \vv^\delta + \tfrac{1}{\rho_0} \nabla p^\delta - \nu(\delta)
    \Laplace \vv^\delta - \nu'(\delta) \nabla \Div \vv^\delta 
    &= \vf, \quad \text{in }\Omega^\delta, \label{eq:Navier_Stokes:C}
    \\ 
    - \imath \omega p^\delta + \rho_0 c^2 \Div \vv^\delta 
    &= 0, \quad \text{in }\Omega^\delta, \label{eq:Navier_Stokes:M} \\
    \vv^\delta
    &= \zerobf , \quad \text{on } \partial
      \Omega^\delta, \label{eq:Navier_Stokes:B}\\
    \intertext{%
    with the acoustic velocity $\vv^\delta$, the acoustic pressure
    $p^\delta$, the mean density $\rho_0 > 0$,
    the speed of sound $c$, the kinematic and secondary viscosities
    $\nu(\delta), \nu'(\delta) > 0$. %
    We scale the viscosities for $\delta \to 0$ like $\delta^4$ such that the size of the viscous boundary layers 
    remain asymptotically the same at the scale of a single hole. %
    If the duct is modelled to be of infinite extend then additional conditions at infinity have to be posed, \eg,
    for a channel of constant cross section with infinite extend in $z\pm\infty$ these conditions are 
    }
    \lim_{z \to \pm \infty} p^\delta 
    &=0. \label{eq:Navier_Stokes:I}
  \end{align}
\end{subequations}
Moreover, we assume the souce to be located away from the perforated wall such that $\vf = 0$ in a neighbourhood.

In the following section we study the solution of the viscoacoustic
model in three different geometrical scales beginning at the scale of
one hole, pursuing with the scale of one period of the hole array and
concluding with the macroscopic scale on which the impedance
conditions follow.

\subsection{\sffamily Microscopic scale: the near field around one hole}
\label{sec:near-field-around}

\begin{figure}[!bt]
  \centering
  \includegraphics[width=0.8\linewidth]{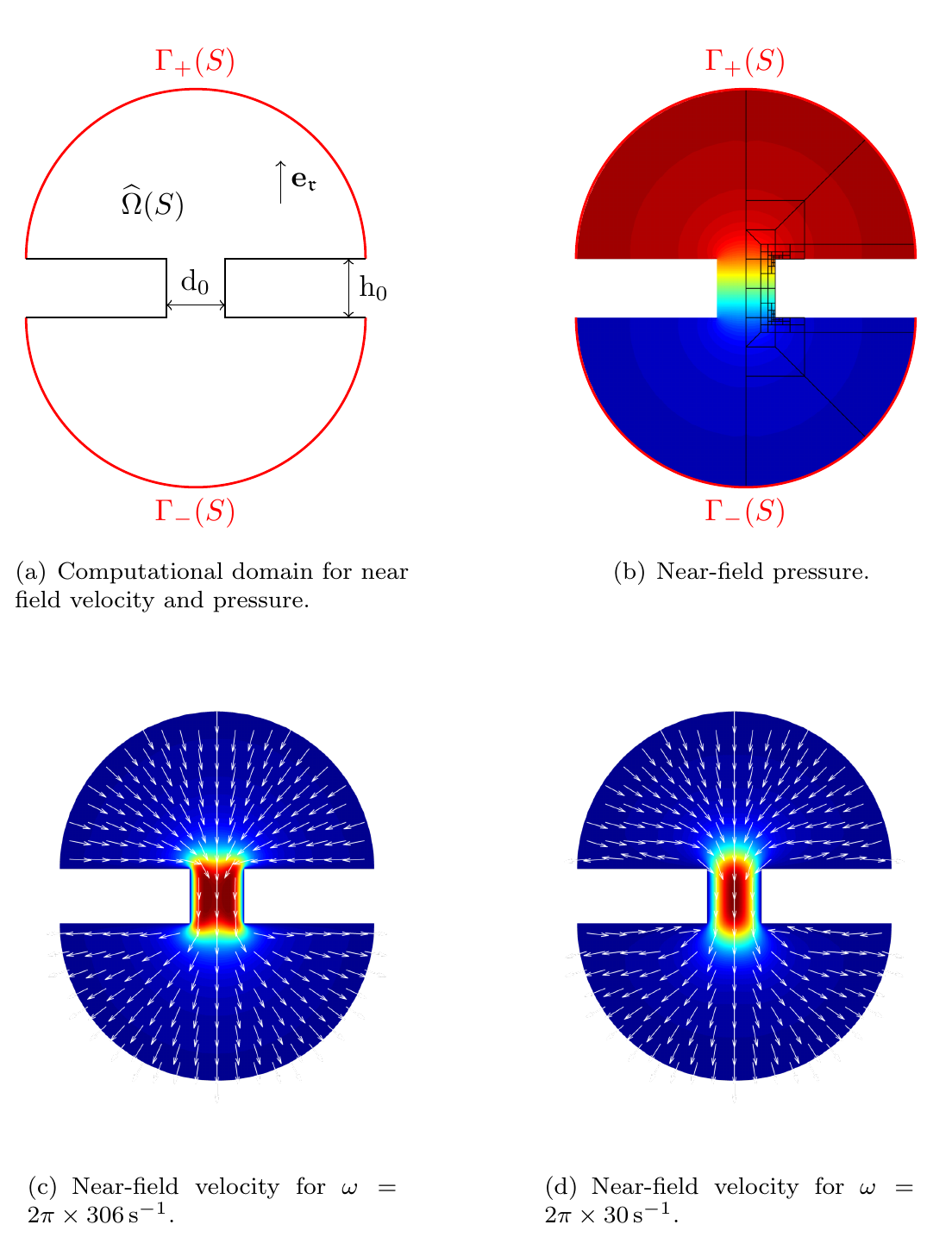}
  \caption{(a) Computational domain for the near field problem around
    a single hole. (b) The near field pressure (real part) for the
    liner configuration DC006 (see
    Table~\ref{tab:liner_configurations}) at
    $\omega = 2\pi\times 306 \,\text{s}^{-1}$ using $S = 40$.  (c),(d)
    The near field velocity (imaginary part) for the same
    configuration as (b) and for
    $\omega=2\pi \times 30\,\text{s}^{-1}$.  Here, the color coding
    corresponds to the amplitude and the arrows to the direction of
    the velocity.  }
  \label{fig:NNF}
\end{figure}

In the vicinity of one hole that tends to a point $\vx_\Gamma$ on the
interface $\Gamma_\liner$ %
we use the local coordinate
$\fX := (\fr,\fy,\fz) =
((r-\Rd)/\delta^2,r\theta/\delta^2,z/\delta^2)$. %
As $\delta \to 0$, the hole variable $\fX$ occupies the whole
unbounded domain $\widehat{\Omega}$ defined by (see
Fig.~\ref{fig:NNF}a)
\begin{equation}
  \label{eq:Omegahat}
  \widehat{\Omega} = \big\lbrace (\fr,\fy,\fz) \in \IR^3 \text{ such that
  } \fr < 0 \text{ or } \fr > \rh_0 \rbrace \cup \widehat{\Omega}_{\hole} \ 
\end{equation}
where $\widehat{\Omega}_{\hole}$ is the scaled domain representing one
hole, and we assume $\zerobf \in \widehat{\Omega}_{\hole}$.  For
instance a vertical cylindrical hole of diameter $\rd_0\delta^2$ can
be represented by
$\widehat{\Omega}_{\hole} = \{ (\fr,\fy,\fz) \in \IR^3 \text{ such
  that } 0 \leq \fr \leq \rh_0 \text{ and } \fy^2 + \fz^2 < \frac14
\rd_0^2 \}$.
  
Close to one hole of the perforated liner, 
we represent the solution $(\vv^\delta,p^\delta)$ of~\eqref{eq:Navier_Stokes}~as
\begin{equation}
  \label{eq:expansion_NNF}
  \begin{aligned}
    \vv^\delta & = \vfv_0(\vx_\Gamma,\fX) +O(\delta)\ , \\
    p^\delta & = \fp_0(\vx_\Gamma,\fX) + \delta^2
    \fp_1(\vx_\Gamma,\fX) + O(\delta^3)\ ,
  \end{aligned}
\end{equation}
where the near field corrector terms $(\vfv_0,\fp_0,\fp_1)$ do not depend on
$\delta$. The scaling of the second corrector for the pressure as
$\delta^2$ is due to the associated scaling of the size of the holes.

%
Now, inserting expansion~\eqref{eq:expansion_NNF} into the
viscoacoustic model~\eqref{eq:Navier_Stokes} and identifying formally
terms of same powers of $\delta$ results first in the fact that
$\fp_0(\vx_\Gamma,\fX)$ is a constant function in $\fX$, and
then in a product representation of the
near-field corrector
\begin{equation*}
  \left(\vfv_0(\vx_\Gamma,\fX),\fp_1(\vx_\Gamma,\fX)\right) =
  c(\vx_\Gamma) \left(\tilde{\vfv}(\fX),\tilde{\fp}(\fX)\right)\ ,
\end{equation*}
where $c(\vx_\Gamma)$ allows for a slow variation of near field velocity and pressure along the wall.

The near field profiles $(\tilde{\vfv},\tilde{\fp})$ are solution of
the instationary Stokes problem
\begin{subequations}
\label{eq:canonical_NNF}
\begin{align}
  - \imath \omega \tilde{\vfv} + \tfrac{1}{\rho_0}\nabla \tilde{\fp} - \nu_0 \Laplace
  \tilde{\vfv} & = \zerobf, && \quad \text{in } \widehat{\Omega}, \\
  \label{eq:canonical_NNF:2}
  \Div \tilde{\vfv} & = 0, && \quad \text{in } \widehat{\Omega}, \\
  \tilde{\vfv} & = \zerobf , && \quad \text{on } \partial \widehat{\Omega},
  \label{eq:canonical_NNF:3}
  \intertext{where $\nabla$, $\Div$ and $\Delta$ are the gradient, divergence and Laplace operator in $\fX$
  (cf. \cite[Sec.~2.1.6]{PopieDiss} in time-domain).
   The near field velocity profile is incompressible on the scale of one hole
   and fulfills together with the near field pressure profile
   the Stokes equations with an at the scale of one hole significant viscosity $\nu_0$ 
   and the additional term $-\imag\omega\tilde{\vfv}$ that reflects a time shift between excitation and excited fields. %
   These equations are completed by Dirichlet jump conditions at infinity 
   }
  \label{eq:canonical_NNF_infinity}
  \lim_{S \to \infty} \tilde{\fp}_{|\Gamma_\pm(S)} &= \pm \tfrac{1}{2}\ ,
\end{align}
\end{subequations}
that act as a excitation from far away and will be used for the
matching with the mesoscopic scale (see Sec.~\ref{sec:mesoscopic
  scale}). Here,
\begin{equation}
  \label{eq:SphericalBoundaries}
    \Gamma_\pm(S) = \left\lbrace (\fr,\fy,\fz) \in \widehat{\Omega},
      \pm \fr > \fr_\pm \text{ and } (\fr - \fr_\pm)^2 + \fy^2 + \fz^2 = S^2
    \right\rbrace,
\end{equation}
with $\fr_-=0$ and $\fr_+ = \rh_0$, are the two half-spheres 
(see Fig.~\ref{fig:NNF}a) that are moved towards infinity.

Note that in problem~\eqref{eq:canonical_NNF} the term
$-\nu_0' \nabla \Div \tilde{\vfv}$ that would appear in the first line
cancels out due to the divergence free condition~\eqref{eq:canonical_NNF:2}. 
Moreover, note that the term
$-\nu_0 \Delta \tilde{\vfv}$ can be replaced by $\nu_0 \Curl\Curl \tilde{\vfv}$
and so only
the vorticity part of the velocity $\tilde{\vfv}$ will exhibit a viscosity boundary layer as we will see later.

Problem~\eqref{eq:canonical_NNF}
is a classical saddle-point problem and admits a unique solution
stated by the following
\begin{prop}
  There exists a unique solution
  $(\tilde{\vfv},\tilde{\fp}) \in (\Hone(\widehat{\Omega}))^3 \times
  \fnV(\widehat{\Omega})$
  of~\eqref{eq:canonical_NNF}, where
  $\fnV(\widehat{\Omega}) = \left\lbrace P \in
    \Honeloc(\widehat{\Omega}) \text{ such that } \xnorm{\nabla
      P}{\Ltwo(\widehat{\Omega})} < \infty \right\rbrace.$
\end{prop}
Note, that the pressure space $\fnV(\widehat{\Omega})$ allows for a constant behavior towards infinity.

With the near field velocity profile $\tilde{\vfv}$ defined
by~\eqref{eq:canonical_NNF} we can define in analogy to the Rayleigh
conductivity {\it a posteriori} the quantity
\begin{equation}
  \label{eq:KR}
  k_R := \lim_{S \to \infty} \frac{\imath \omega \rho_0}{2} \Big( \int_{\Gamma_+(S)}
  \tilde{\vfv} \cdot \vn - \int_{\Gamma_-(S)}
  \tilde{\vfv} \cdot \vn \Big)
\end{equation}
using the volume flux towards infinity in a symmetric
way. Here, $\vn$ is the outer normal vector. %
In this way, the quantity $k_R$ is a mapping of a constant near field
pressure at infinity to the flux at infinity.  To see the analogy it
suffices to consider time harmonic fields varying like
$\exp(-\imath \omega t)$, the volume flux $Q(t)$ through the aperture
counted positively along the direction of the $\ve_\fr$ axis to be the
same as the volume flux through the surface $\Gamma_+(S)$
(respectively $\Gamma_-(S)$), counted positively (resp. negatively)
along the direction of the normal vector $\vn$, and to
compare~\eqref{eq:KR_literature} and~\eqref{eq:KR}.

Note, that the normal component of the near field velocity profile $\vfv$ decays
like $1/S^2$ towards infinity and combines different behaviour close
to and away from the wall (see Fig.~\ref{fig:NNF}(c) and (d)).  This
behaviour can be rigorously justified with similar techniques as
in~\cite{Nazarov2008,Semin.Delourme.Schmidt:2017}.

For the usual definition of the Rayleigh conductivity $K_R$ it is not
evident where the difference of the pressure -- as it varies locally --
and the volume flux -- as in the original acoustic equations the fluid is
compressible -- shall be evaluated. The quantity $k_R$ is, however,
clearly defined by~\eqref{eq:canonical_NNF} and~\eqref{eq:KR} as the
near field pressure tends to constant values for $|\fX| \to \infty$ and
as the near field velocity is incompressible. %
This results from the separation of the effects at the different
length scales, namely viscous incompressible behaviour in the vicinity
of the holes versus inviscid, compressible behaviour away from them,
due to the asymptotic ansatz. %
As the near field profiles are defined in local coordinates $\fX$ it
has the dimensions of one over length and we denote it as effective
Rayleigh conductivity of the liner. %

The definition of the effective Rayleigh conductivity $k_R$ can be
used for inviscid fluids as well for which $\nu_0 = 0$ if the no-slip
boundary conditions~\eqref{eq:canonical_NNF:3} are replaced by
$\vfv\cdot\vn = 0$.

\subsection{\sffamily Mesoscopic scale: the hole pattern}
\label{sec:mesoscopic scale}

\begin{figure}[!bt]
  \centering
  \includegraphics[width=0.75\linewidth]{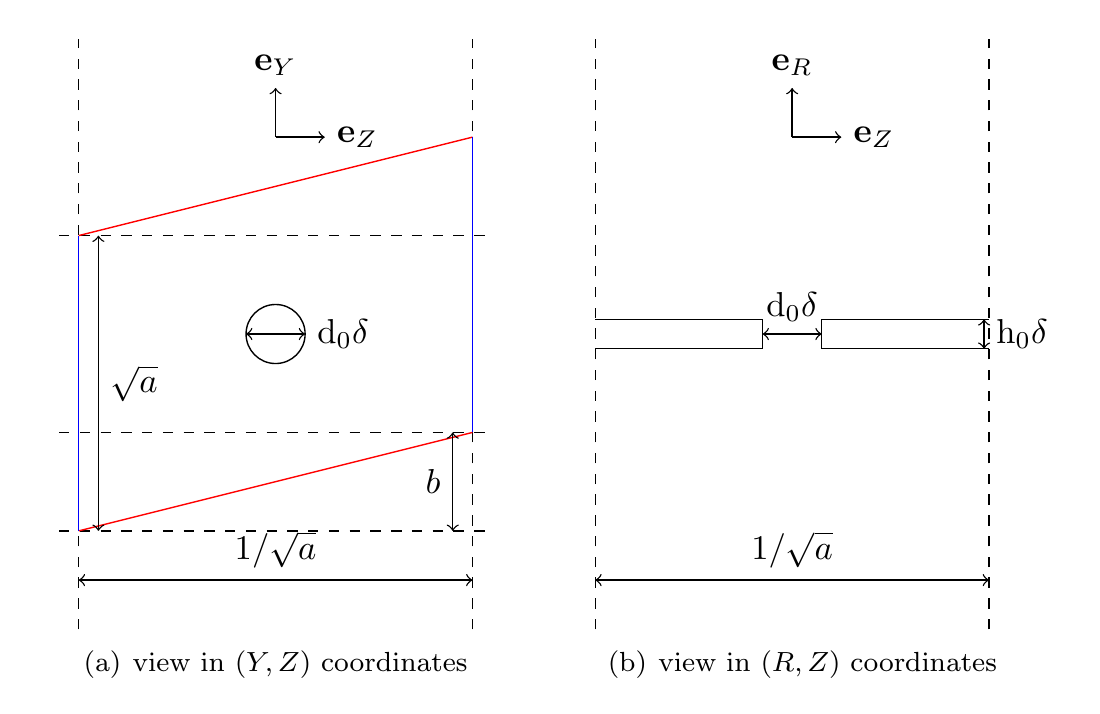}
  \caption{Representation of the periodicity cell $\nB^\delta$ associated
    to the intermediate scale}
  \label{fig:intermediate_scale}
\end{figure}

Pursuing with the scale of one period of the hole array and in the
vicinity of one hole that tends to $\vx_\Gamma$, we use the local
coordinate
$\vX := (R,Y,Z) = ((r-\Rd)/\delta,r\theta/\delta,z/\delta)$.  We
consider for fixed $\delta > 0$ the infinite periodicity cell
\begin{equation}
  \label{eq:Bdelta}
  \begin{aligned}
    \nB^\delta &= \nB^\delta_+ \cup \nB^\delta_-
    \cup \delta \widehat{\Omega}_\hole
  \end{aligned}
\end{equation}
where
$\nB^\delta_\pm = \big\lbrace (R,Y,Z) \in \IR^3 \text{ such that } |Y
- bZ| < \tfrac{\sqrt{\ra}}{2}, |Z| < \tfrac{1}{2\sqrt{a}}, \pm R >
R_\pm^\delta\big\rbrace$
with $R_-^\delta = 0$, $R_+^\delta = \rh_0\delta$ are two
semi-infinite parallelepipeds whose opposite lateral faces
$|Z| = \tfrac{1}{2\sqrt{a}}$ and $|Y - bZ| = \tfrac{\sqrt{\ra}}{2}$ are considered to be
identified with each other such that $\nB^\delta_\pm$ and so
$\nB^\delta$ are topologically equivalent to a torus.  With the
cross-section of the periodicity cell
\begin{align*}
  \Gamma(S) = \{(R,Y,Z) \in \IR^3 \text{ such that } |Y - bZ| < \tfrac{\sqrt{\ra}}{2},|Z| < \tfrac{1}{2\sqrt{a}}, R = S\}
\end{align*}
the symmetric difference $A \triangle B := (A \cup B) \setminus (A \cap B)$
the boundary of the periodicity cell is given as
$
\partial\nB^\delta = (\Gamma(h_0\delta) \cup \Gamma(0))
\,\triangle\, \delta \partial\widehat{\Omega}_\hole$.
It consists of the wall boundary and the boundary of the hole.  The
periodicity cell $\nB^\delta$ degenerates as $\delta \to 0$ and tends
to the union $\nB^0$ of two semi-infinite parallelepipeds $\nB^0_\pm$
connected by the point~$\zerobf$, an infinitely small hole.

Inside the periodic array of holes, we represent the solution
$(\vv^\delta,p^\delta)$ of~\eqref{eq:Navier_Stokes} as
\begin{equation}
  \label{eq:expansion_NF}
  \begin{aligned}
    \vv^\delta & = \vV_0^\delta(\vx_\Gamma,\vX)+ O(\delta)\ , \\
    p^\delta & = P_0^\delta(\vx_\Gamma,\vX) + \delta
    P_1^\delta(\vx_\Gamma,\vX) + O(\delta^2) 
  \end{aligned}
\end{equation}
with $\vX \in \nB^\delta$.

%

Inserting expansion~\eqref{eq:expansion_NF} in
problem~\eqref{eq:Navier_Stokes} and 
identifying formally the terms of same
powers of $\delta$ gives that $P_0^\delta(\vx_\Gamma,\vX)$ is constant in $\vX$
and a separation of variables for the mesoscopic corrector as 
$\big(\vV_0^\delta(\vx_\Gamma,\vX),P_1^\delta(\vx_\Gamma,\vX)\big) = c(\vx_\Gamma) \big(\fnV^\delta(\vX),\nP^\delta(\vX)\big)$
with the mesoscopic profile $(\fnV^\delta$, $\nP^\delta)$ satisfying the Darcy-type problem
\begin{equation}
  \label{eq:nV_nD_delta}
  \left\lbrace
    \begin{aligned}
      - \imath \omega \fnV^\delta + \tfrac{1}{\rho_0} \nabla \nP^\delta & = 0, \quad
      \text{in }\nB^\delta,\\
      \Div \fnV^\delta & = 0, \quad \text{in }\nB^\delta, \\
      \fnV^\delta \cdot \vn & = 0, \quad \text{on } \partial\nB^\delta\ .
    \end{aligned}
  \right.
\end{equation}
Here, $\nabla$ and $\Div$ are the gradient and divergence in $\fX$.
The formal identification of terms of same power in $\delta$ can be justified 
despite the fact that the size of the hole depends on $\delta$ as well. For this
an additional scale $\eta$ for the size of one hole has to be introduced that is first considered to be independent of $\delta$ 
due to its different meaning and later fixed to $\delta^2$. The expansion~\eqref{eq:expansion_NF} is then in $\delta$, where the terms of the expansion depend on $\eta$.
For the brevity of the article we have chosen directly $\eta = \delta^2$.

Note that~\eqref{eq:nV_nD_delta} is equivalent to an homogeneous
Laplace problem with Neumann boundary conditions for the pressure
profile $\nP^\delta$, where the velocity profile $\fnV^\delta$ can be
computed from.
Following~\cite[Proposition~2.2]{Delourme.Schmidt.Semin:2016}, we can
therefore  state the following
\begin{prop}
  \label{prop:meso_uniqueness}
  For any fixed $\delta > 0$, the kernel of
  problem~\eqref{eq:nV_nD_delta} is of dimension~$2$ and spanned by
  the functions $(\fnV_\nN^\delta,\nP_\nD^\delta) = (0,1)$ and
  $(\fnV_\nD^\delta,\nP_\nD^\delta)$ such that $\nD_v^\delta$ is constant
  as $R \to \pm \infty$. Moreover, there exists
  $\nD_\infty^\delta \in \IC$ such that $(\fnV_\nD^\delta,\nP_\nD^\delta)$
  admits the following limit behaviour:
  \begin{equation}
    \label{eq:nV_nD_infty_delta}
    \begin{aligned}
      \fnV_\nD^\delta &= \tfrac{1}{\imath \omega} \ve_R +o(1),&& \quad R
      \to \pm \infty,\\
      \nP_\nD^\delta & = \rho_0 R \pm \nD_\infty^\delta + o(1),&& \quad R
      \to \pm \infty.
    \end{aligned}
  \end{equation}
\end{prop}
It remains to determine the constant $\nD_\infty^\delta$, where we are
in particular interested in its asymptotic behaviour for $\delta\to0$.
To obtain this behaviour we will match the mesoscopic functions
$\fnV_\nD^\delta$ and $\nP_\nD^\delta$ with the near field profiles
$\tilde{\vfv}$ and $\tilde{\fp}$ at half-spheres
$\Gamma_\pm(s^\delta)$ of radius~$s^\delta$
for $\sqrt{\delta} < s^\delta < 2\sqrt{\delta}$ centered at the aperture $\zerobf$.  First we note
that due to the incompressibility and the limit behaviour of
$\fnV_\nD^\delta$ for its volume flux over the half-spheres it holds
\begin{equation*}
  \frac{\imath\omega}{2} \left(\int_{\Gamma_+(s^\delta)}
    \fnV_\nD^\delta \cdot \vn - \int_{\Gamma_-(s^\delta)}
    \fnV_\nD^\delta \cdot \vn\right) 
  = \frac{\imath \omega}{2} \lim_{S\to \infty} \int_{\Gamma(S)}
  \fnV_\nD^\delta \cdot \ve_R + \int_{\Gamma(-S)} \fnV_\nD^\delta
  \cdot \ve_R = 1\ .
\end{equation*}
Using this equality, definition~\eqref{eq:KR} of the effective
Rayleigh conductivity $k_R$, the mesoscopic to microscopic variable
change $\fX = \vX/\delta$, and matching of the mesoscopic velocity
$\fnV_\nD^\delta$ and the near field velocity profile $\tilde{\vfv}$
we find that
\begin{equation*}
  \fnV_\nD^\delta(\vX) \sim \frac{\rho_0}{k_R \delta^2}
  \tilde{\vfv}(\tfrac{\vX}{\delta}) \quad
  \text{ for } \sqrt{\delta} < |\vX| < 2\sqrt{\delta} \text{ and } \delta \to 0\ .
\end{equation*}
By linearity and using definition of
  problems~\eqref{eq:canonical_NNF} and~\eqref{eq:nV_nD_delta}, the
  gradient of the mesoscopic pressure $\nP_\nD^\delta$ can be matched
  with the gradient of the near field pressure profile as
  well. Integrating these gradients, using
  limit~\eqref{eq:canonical_NNF_infinity} and
  Proposition~\ref{prop:meso_uniqueness} leads to
\begin{equation*}
  \nP_\nD^\delta(\vX) \sim \frac{\rho_0}{k_R\delta}
  \tilde{\fp}(\tfrac{\vX}{\delta}) \sim \pm\frac{\rho_0}{2 k_R\delta}
  \quad \text{ for } \sqrt{\delta}< |\vX| < 2\sqrt{\delta}, \pm R > 0 \text{ and }
  \delta \to 0\ . 
\end{equation*}
As for $\delta \to 0$ the mesoscopic pressure $\nP_\nD^\delta$ tends to $\rho_0 R \pm \nD^\delta_\infty$ 
if $\delta = o(|\vX|)$ we conclude that
\begin{align}
  \label{eq:jump_towards_infinity}
  \nD_\infty^\delta = \frac{\rho_0}{2 k_R\delta} + o(\delta^{-1})\ .
\end{align}
This blow up of the coefficient $\nD_\infty^\delta$ as $\delta \to 0$ 
in accordance with its numerical computations based on an asymptotic analysis of~\eqref{eq:Navier_Stokes} with only two scales~\cite{Semin.Schmidt:2017},
where the hole size is considered not as a scale but as a parameter.

\subsection{\sffamily Macroscopic scale and impedance conditions}

Finally, away from a vicinity of the layer, the solution
$(\vv^\delta,p^\delta)$ of~\eqref{eq:Navier_Stokes} is represented~by
\begin{equation}
  \label{eq:expansion_FF}
  \begin{aligned}
    \vv^\delta(\vx) & = \vv_0(\vx) + o(1), \\
    p^\delta(\vx) & = p_0(\vx) + o(1).
  \end{aligned}
\end{equation}
Inserting this expansion in problem~\eqref{eq:Navier_Stokes} and
making a formal identification in terms of powers of $\delta$ gives
that $(\vv_0,p_0)$ is solution of the classical Helmholtz problem
\begin{subequations}
  \label{eq:Helmholtz}
  \begin{align}
    \label{eq:Helmgoltz:C}
    - \imath \omega \vv_0 + \tfrac{1}{\rho_0}\nabla p_0 &=\vf, \quad \text{in }\Omega
    \setminus \Gamma_\liner, \\
    - \imath \omega p_0 + \rho_0 c^2 \Div \vv_0 &=0, \quad \text{in }\Omega
    \setminus \Gamma_\liner,
    \intertext{and a multiscale analysis~\cite{Schmidt.Thoens.Joly:2014} for rigid walls leads to the boundary conditions}
    \label{eq:Helmgoltz:B}
    \vv_0 \cdot \vn &=0, \quad \text{on }\partial\Omega\ .
  \end{align}
  
  The limit condition~\eqref{eq:Navier_Stokes:I} becomes
  \begin{equation}
    \label{eq:Helmholtz:I}
    \lim_{z \to \pm \infty}  \vv_0 \cdot (\pm 1,0,0)
    - T_\pm^1 p_0 = 0,
  \end{equation}
  \end{subequations}  
  where $T_\pm^1$ is the Dirichlet-to-Neumann operator based on the
  projection on the outgoing propagative modes,
  see~\cite[Eq.~(2.7)]{Goldstein:1982}
  and~\cite{Semin.Schmidt:2016}. %
  This problem is completed by jump conditions across the interface
  $\Gamma_\liner$. %
  To obtain the conditions we match the macroscopic pressure $p_0$
    and flux $\vv_0 \cdot\vn$ in a matching zone at distance
    $\sqrt{\delta}$ to the interface $\Gamma_\liner$ to the mesoscopic
    pressure and velocity functions. %
    For the pressure we find
  \begin{multline}
     \label{eq:p0_matching}
     p_0(\vx)  = C_\nN(\vx_\Gamma) \, \nP_\nN^\delta(\tfrac{\vx - \vx_\Gamma}{\delta}) %
     + \delta C_\nD(\vx_\Gamma) \, \nP_\nD^\delta(\tfrac{\vx - \vx_\Gamma}{\delta}) \\
     \text{ for } \sqrt{\delta} \leq |\vx - \vx_\Gamma| \leq 2\sqrt{\delta}  \text{ and } \delta \to 0.
  \end{multline}
  with two functions $C_\nN$, $C_\nD$ that allow for slow variation along the perforated wall. %
  With the factor $\delta$ the limit $\delta \nP^\delta_\nD(\frac{\vx - \vx_\Gamma}{\delta})$ for $\delta\to0$ remains bounded.
  Subtracting the two limits of~\eqref{eq:p0_matching} for $\delta \to 0$ we obtain
  \begin{align}
    \label{eq:jump_p0_CD}
     \bjump{p_0}(\vx_\Gamma) &:= %
     \lim_{\delta\to0} p_0(\vx_\Gamma + \sqrt{\delta} \vn) - p_0(\vx_\Gamma - \sqrt{\delta} \vn)
     = C_\nD(\vx_\Gamma) \frac{\rho_0}{k_R}\ .
  \end{align}
  Taking the gradient in $\vx$ on both sides of~\eqref{eq:p0_matching} and using~\eqref{eq:Helmgoltz:C}, the assumption that $\vf = 0$ close to the perforated wall
  and~\eqref{eq:nV_nD_delta} we find
  \begin{multline}
    \label{eq:v0_matching}
    \vv_0(\vx)\cdot\vn = \frac{\rho_0}{\imath\omega}\nabla p_0(\vx)\cdot\vn
    = \delta C_\nD(\vx_\Gamma) \, \frac{\rho_0}{\imath\omega} \nabla \nP_\nD^\delta(\tfrac{\vx - \vx_\Gamma}{\delta})
    = C_\nD(\vx_\Gamma) \, \fnV_\nD^\delta(\tfrac{\vx - \vx_\Gamma}{\delta})\\
    \text{ for } \sqrt{\delta} \leq |\vx - \vx_\Gamma| \leq 2\sqrt{\delta}  \text{ and } \delta \to 0\ .
  \end{multline}
  As the two limits for $\fnV^\delta_\nD$ for $R\to\pm\infty$ coincide we obtain
  \begin{subequations}
    \label{eq:impedance_conditions}
  \begin{align}
    \label{eq:impedance_conditions:velocity}
    \bjump{\vv_0 \cdot \vn}(\vx_\Gamma) &= 0, \quad \text{on }\Gamma_\liner\ .
    \intertext{
  Finally, taking the average of~\eqref{eq:v0_matching} and the limit $\delta\to0$ gives in view of~\eqref{eq:jump_p0_CD} 
  the impedance conditions
    }
    \label{eq:impedance_conditions:pressure}
    \bjump{p_0}(\vx_\Gamma) &= \frac{\imath \omega \rho_0}{k_R} \bavrg{\vv_0
    \cdot \vn}(\vx_\Gamma), \quad \text{on }\Gamma_\liner\ .
  \end{align}
\end{subequations}
Note, that the impedance conditions do not depend on the pattern of
the holes, more precisely on the values $\ra$ and $\rb$ (see
Fig.~\ref{fig:liner_geometry}), but only on their area $A_\delta$,
namely through $\nu_0 = \nu / A_\delta^2$ in the computation of the
effective Rayleigh conductivity~$k_R$.

\paragraph{\sffamily Distinguished limit} Note, that the nature of the impedance
condition~\eqref{eq:impedance_conditions:pressure} is due to the
choice of asymptotic scales. %
It represents a distinguished limit meaning that different choice
would lead to one of the trivial conditions $\bjump{p_0}(\vx_\Gamma) = 0$
(transparent wall) or $\bavrg{\vv_0\cdot\vn}(\vx_\Gamma) = 0$ (rigid
wall).  If we would scale the diameter of each hole like
$\eps(\delta)$ as well as the thickness of the perforated wall such
that $\delta^2 = o(\eps(\delta))$ then we would obtain transparent
wall conditions in the limit $\delta\to0$. \emph{A contrario},
the impedance conditions become rigid wall conditions if we would use
the scaling $\eps(\delta) = o(\delta^2)$.  The choice of asymptotic
scales was already stated in~\cite{Sanchez.Sanchez:1982} for
infinitely thin perforated wall and the Stokes flows.

\paragraph{\sffamily Acoustic Impedance} The nature of the impedance conditions is known in the literature: the
notion of impedance can be found in the works of Webster in the
1910s~\cite{webster1919acoustical}. More precisely, he defines the
normalized specified acoustic impedance $\zeta$ by (note there is a
complex conjugate and a different sign due to the different choice of
the time-dependency convention)
\begin{equation}
  \label{eq:impedance}
  \zeta := - \frac{\overline{\bjump{p_0}}}{c \rho_0 \overline{\bavrg{\vv_0
        \cdot \vn}}}\ .
\end{equation}
For the derived impedance
conditions~\eqref{eq:impedance_conditions:pressure} and by
identification, the normalized specified acoustic impedance for
perforated walls is given by
\begin{equation}
  \label{eq:impedance_our_model}
  \zeta = \frac{\imath \omega }{c \overline{k_R}} = \frac{\imath
    \omega k_R}{c |k_R|^2}\ .
\end{equation}
The resistance $\Re(\zeta)$ and the reactance $\Im(\zeta)$ are
positive quantities when $k_R$ has a positive real part and a negative
imaginary part. Moreover in the inviscid case $k_R$ is a positive real
number, so that the normalized specified acoustic impedance $\zeta$ is
purely a reactance.

\paragraph{\sffamily Formulation in pressure only} 
One can also remark that Problem~\eqref{eq:Helmholtz} can be
formulated in terms of pressure only:
equations~\eqref{eq:Helmgoltz:C}-- \eqref{eq:Helmholtz:I}
give
\begin{subequations}
  \begin{equation}
    \label{eq:Helmholtz:pressure}
    \begin{aligned}
    \Delta p_0 + \tfrac{\omega^2}{c^2} p_0 &= \Div \vf, && \text{in
    }\Omega \setminus \Gamma_\liner, \\
    \nabla p_0 \cdot \vn &= 0,
    && \text{on }\partial \Omega,\\
    \lim_{z \to \pm \infty}  \pm\partial_z p_0 - \imath\omega\rho_0 T_\pm^1 p_0 &= 0,
    \end{aligned}
  \end{equation}
  and impedance
  conditions~\eqref{eq:impedance_conditions:velocity}-%
  \eqref{eq:impedance_conditions:pressure}
  are written in terms of the pressure as
  \begin{equation}
    \label{eq:impedance_conditions:pressure_only}
    \bjump{\nabla p_0 \cdot \vn}(\vx_\Gamma) = 0 \quad \text{and} \quad
    \bavrg{\nabla p_0
        \cdot \vn}(\vx_\Gamma) = k_R \bjump{p_0}(\vx_\Gamma)\ .
  \end{equation}
  This kind of conditions were proposed for the inviscid
  case~\cite{Bendali.Fares.Piot.Tordeux:2012}, where $k_R$ turns out
  to be the effective plate compliance.
\end{subequations}

\section{\sffamily Results and discussion}

In this section, we are interested by the numerical computation of the
effective Rayleigh conductivity $k_R$, the computation of dissipation
losses in acoustic ducts with the impedance conditions and comparison
with data from experimental measurements.

\subsection{\sffamily Numerical computation of $k_R$} 

The effective Rayleigh conductivity $k_R$ is defined through the
solution of the near field velocity and pressure profiles in the
unbounded domain $\widehat{\Omega}$ around a single hole. To compute
$k_R$ numerically we truncate the unbounded domain, on which we use
the finite element method for discretization and propose an
extrapolation procedure to increase the accuracy.

First, we define the truncated domain 
\begin{equation}
  \label{eq:widehat_Omega_set}
  \widehat{\Omega}(S) = \widehat{\Omega} \cap \big( \max(|\fX -
  (0,0,0)|, |\fX - (\rh_0,0,0)|) < S \big)
\end{equation}
of $\widehat{\Omega}$ for a given truncation radius
$S > 0.5 \sqrt{\rd_0^2+\rh_0^2}$ (see Fig.~\ref{fig:NNF}(a)). It has
two artificial boundaries $\Gamma_\pm(S)$ that are no boundaries of
$\widehat{\Omega}$. We restrict the problem~\eqref{eq:canonical_NNF}
to $\widehat{\Omega}(S)$ and
$\partial \widehat{\Omega}(S) \cap \partial \widehat{\Omega}$, and we
approximate the conditions~\eqref{eq:canonical_NNF_infinity} by setting
\begin{equation*}
  \tilde{\fp}_{|\Gamma_\pm(S)} = \pm \tfrac{1}{2}.
\end{equation*}

\begin{figure}[!bt]
  \centering
  \includegraphics[width=0.65\linewidth]{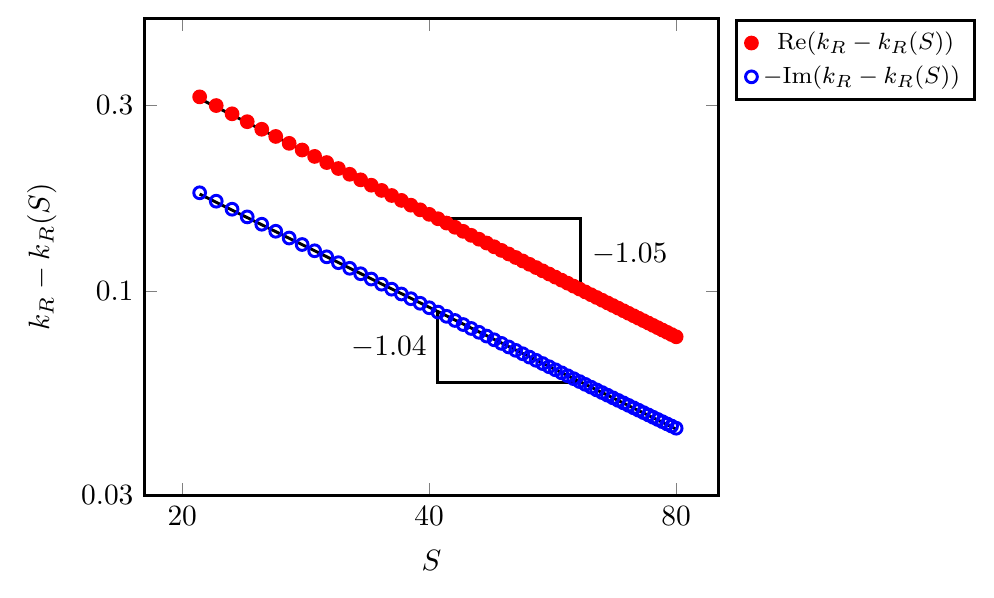}
  \caption{Convergence of the real and imaginary parts of the
    approximated Rayleigh conductivity $k_R(S)$ to its limit value
    $k_R = 4.513-1.210\imath$ in dependence of the truncation radius
    $S$ for the liner DC006 at frequency $f = 306\,\text{Hz}$ (see
    Table~\ref{tab:liner_configurations}).}
  \label{fig:GR_vs_S}
\end{figure}

From the resolution of the truncated problem we compute the
approximated Rayleigh conductivity $k_R(S)$ taking as well an
approximation of~\eqref{eq:KR}, namely
\begin{equation}
  \label{eq:KR_truncated}
  k_R(S) := \frac{\imath \omega\rho_0}{2} \Big( \int_{\Gamma_+(S)}
  \tilde{\vfv} \cdot \vn - \int_{\Gamma_-(S)}
  \tilde{\vfv} \cdot \vn \Big),
\end{equation}
Its approximated value $k_R(S)$ tends to the Rayleigh conductivity
$k_R$ as $1/S$ as illustrated in Fig.~\ref{fig:GR_vs_S}. This
first-order convergence can be explained with a rigorous analysis of
the solution of problem~\eqref{eq:canonical_NNF} towards infinity
using the Mellin transform~\cite{Kozlov.Mazya.Rossmann:1997} and
showing that the solution of this problem on $\Gamma_\pm(S)$ is a
superposition of a radial expansion with respect to $1/S$ and of a
cartesian expansion with terms decaying exponentially with respect to
the distance to the boundary. %
Similar analyses were performed for the Poisson and Helmholtz problems
in conical domains with a rough periodic boundary~\cite{Nazarov2008}
or perforated wall~\cite{Semin.Delourme.Schmidt:2017}.

As, more precisely, the Rayleigh conductivity $k_R$ can be expanded in
powers of~$1/S$ we use an extrapolation in $1/S$ of first order
approximations $k_R(S)$ for different truncation radia $S$ to obtain a
second or higher order approximation of the limit value $k_R$.

For the particular case of a straight cylindrical hole that is without
loss of generality centered at $\fy=\fz=0$, the domain
$\widehat{\Omega}(S)$ is invariant under rotation around the $\fr$
axis as well as the solution of the problem~\eqref{eq:canonical_NNF}
for the near field profiles.  Hence, the finite element method in two
dimensions can be used for the numerical resolution in a 2D
axis-symmetry setting.  To resolve the boundary layer of size
$\sqrt{\nu_0 / \omega}$ on the wall boundary (cf.~\cite[Sec.~3.1]{PopieDiss}) we use the $hp$-adaptive
strategy of Schwab and Suri~\cite{Schwab.Suri:1996} (see the mesh
shown in Fig.~\ref{fig:NNF}(b)).

\begin{table}[!bt]
  {\scriptsize
    \begin{tabular}{c|c|c|c|c|c|c}
      Config. 
      & number of
      & longitudinal
      & azimuthal
      & hole
      & liner
      & $\sigma$ \\
      & holes (longitudinal, 
      & inter-hole
      & inter-hole
      & diameter
      & thickness
      &  \\
      & azimuthal)
      & distance $\delta / \sqrt{\ra}$ (mm)
      & distance $\sqrt{\ra} \delta$ (mm)
      & $\rd_0 \delta^2$ (mm) 
      & $\rh_0 \delta^2$ (mm) & \% \\
      \hline
      DC006 & (7,52) & 8.5 & 8.45 & 1 & 1 & 1.1 \\
      DC007 & (3,20) & 22 & 21.99 & 2.5 & 1 & 1.0 \\
      DC008 & (7,52) & 8.5 & 8.45 & 2.5 & 1 & 6.8 \\
      DC009 & (3,20) & 22 & 21.99 & 1 & 1 & 0.2 \\
    \end{tabular}
  }
  \caption{Liner configurations. The length of the liner is
    $L=60\,\text{mm}$. The value of the viscosity is 
    $\nu(\delta) = 1.4660 \times 10^{-5} \, \text{m}^2 /
    \text{s}$. For all these configurations $b=0.5\sqrt{\ra}$.}
  \label{tab:liner_configurations}
\end{table}

For four liner configurations, see
Table~\ref{tab:liner_configurations}, from experimental
studies~\cite{Lahiri:2014,lahiri2017a} we have computed the near field
velocity and pressure profiles and so the effective Rayleigh conductivity. %
The relative kinematic viscosity $\nu_0$ is computed as quotient of
the kinematic viscosity
$\nu = 1.4660 \times 10^{-5} \, \text{m}^2 / \text{s}$ of air at
$15^\circ \mathrm{C}$ divided by the period $\delta$ to the power of
four.  In Fig.~\ref{fig:NNF}(b) and Fig.~\ref{fig:NNF}(c) we
illustrate the near field pressure and velocity profiles $\tilde{\fp}$
and $\tilde{\vfv}$ for the liner DC006 at frequency $306\,\text{Hz}$
using a truncation radius $S=40$. %
It is visible that the pressure decays almost linearly inside the
cylindrical hole, but also the behaviour at distance to the
hole. Moreover, the pressure shows close to the rim of the cylinder an
edge singularity (\ie a corner singularity for the 2D axis-symmetric
problem) that is resolved numerically by the {\em hp}-adaptive
refinement strategy. %
The near velocity profile shows a flux from all sides to and through
the hole. It appears that the outward flux of the imaginary part of
$\tilde{\vfv}$ over $\Gamma_+(S)$ is negative (resp. positive over
$\Gamma_-(S)$) corresponding to a positive real part of the
approximate Rayleigh conductivity $k_R(S)$
(see~\eqref{eq:KR_truncated}) and so of the Rayleigh conductivity
$k_R$.  This is in line with the inviscid case, where $k_R$ is real
and positive.  Moreover, we see the higher velocity amplitude inside
the hole that decays towards its boundaries.  This boundary layer
phenomena is more visible for lower frequencies (see
Fig.~\ref{fig:NNF}(d)), where one also see a local change of the
velocity direction on the wall boundary.

\begin{figure}[!bt]
  \centering
  \includegraphics[width=0.95\linewidth]{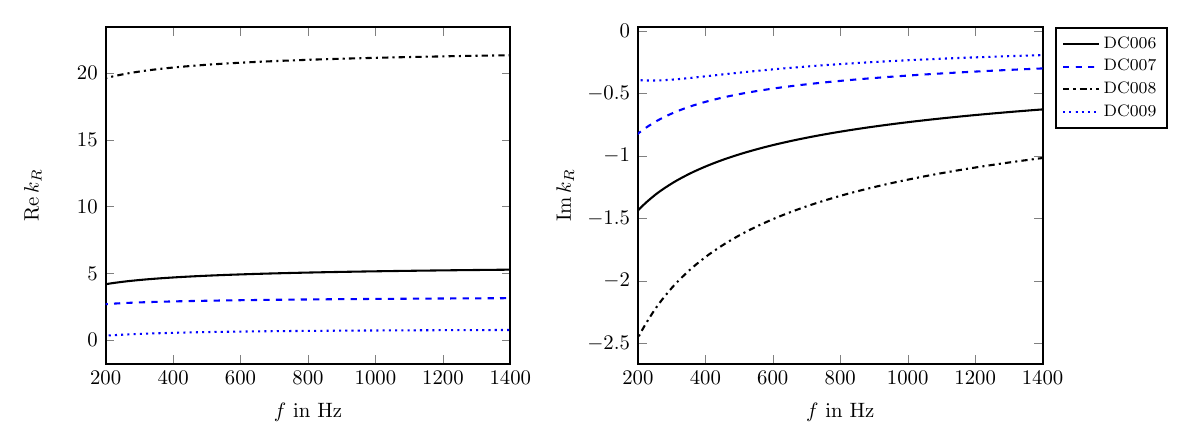}
  \caption{Real and imaginary parts of $k_R$ in dependence of the frequency $f=\tfrac{\omega}{2\pi}$.}
  \label{fig:GR_vs_freq}
\end{figure}

In Fig.~\ref{fig:GR_vs_freq}, we plot the effective Rayleigh
conductivity $k_R$ as a function of the
frequency $f := \tfrac{\omega}{2\pi}$ for different liner
configurations given in
Table~\ref{tab:liner_configurations}. As expected, following the
remark on the normalized specified acoustic impedance~$\zeta$, the
real part of $k_R$ is positive and its imaginary part is negative. One
can also remark that for liner configurations DC006 and DC007, that
have a close value of the porosity $\sigma$ but quite different hole
repartition and hole diameter, their Rayleigh conductivities differ
significantly in both their real and imaginary part.

\begin{figure}[!bt]
  \centering
  \includegraphics[width=0.95\linewidth]{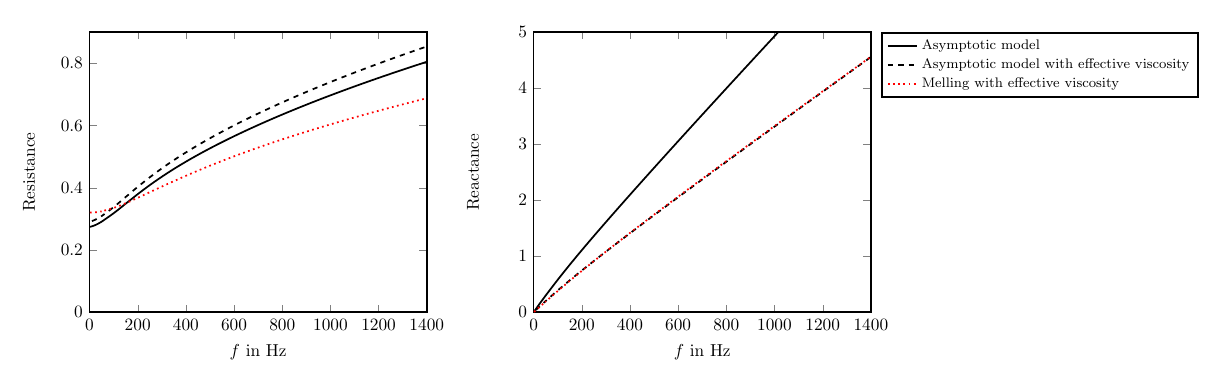}
  \caption[Impedance]{%
    Comparison of the impedance with the normalized specific
    resistance ({\em left}) and the normalized specific reactance
    ({\em right}) computed by our model and Melling model as function
    of the frequency $f=\tfrac{\omega}{2\pi}$ for the liner
    configuration DC006.  As Melling model uses an effective viscosity
    taking into account thermal conductivity losses we show the
    impedance for our model with the effective viscosity as well.  }
  \label{fig:CR_vs_freq}
\end{figure}

In Fig.~\ref{fig:CR_vs_freq}, we show the computed normalized specific
acoustic impedance $\zeta$ for the liner configuration DC006 in
comparison with the Melling model (see~\cite{melling1973acoustic}
and~\cite[Eq.~(12)]{lahiri2017a}), that is given an analytic formula.
For the latter an effective kinematic viscosity
$\tilde{\nu}(\delta) := 2.179 \nu(\delta)$ is used that shall
incorporate also thermal conductivity losses near a highly conducting
wall, see \cite[p.~239]{crandall:1926} and~\cite[p.~62]{Lahiri:2014}.
We plotted the Rayleigh
conductivities obtained from our model with this effective kinematic
viscosity.  The reactance predicted by the two models are very close,
where the resistence differs by up to $20\%$.  The importance of
taking the thermal conductivity losses into account will be seen in
comparison with the measurements and be discussed later in
Sec.~\ref{sec:numerical-results}.

\subsection{\sffamily Dissipation losses in acoustic ducts}

\subsubsection{\sffamily Experimental Setup and Analysis}
\label{sec:experimental-setup}

\begin{figure}[!bt]
  \centering
  \includegraphics[width=0.95\linewidth]{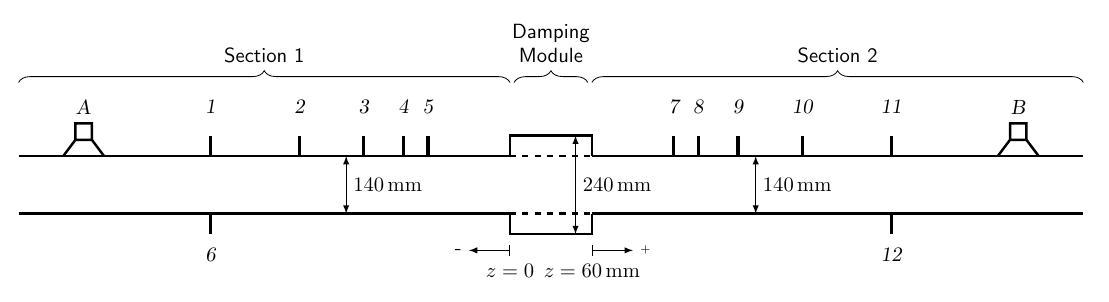}
  \caption{Schematic setup of the Duct Acoustic Testrig (DUCT-C) with
    speakers $A$ and $B$, and microphones 1-12. The anechoic
    terminations at both ends are not shown.}
\label{fig:test-rig}
\end{figure}

The experimental study is performed in the duct acoustic test rig with
a circular cross-section (DUCT-C) at DLR Berlin at ambient
conditions. The setup of the test rig is illustrated in
Fig.~\ref{fig:test-rig}. It allows high precision acoustic
measurements of the damping performance of various liner
configurations, including grazing and bias flow.

The test duct consists of two symmetric measurement sections (section
1 and section 2 in Fig.~\ref{fig:test-rig}) of 1200\,mm length
each. They have a circular cross-section with a radius $\Rd$ of
  70\,mm. In order to minimize the reflection of sound at the end of
the duct back into the measuring section the test duct is equipped
with anechoic terminations at both ends (not shown in
Fig.~\ref{fig:test-rig}). Their specifications follow the ISO 5136
standard. The damping module is a chamber of 60\,mm. It has a
  circular cross-section with a radius of 120\,mm.

A total of 12 microphones are mounted flush with the wall of the test
duct. They are installed at different axial positions upstream and
downstream of the damping module and are distributed exponentially
with a higher density towards the damping module. Two microphones are
installed opposite of each other at the same axial position close to
the signal source. As evanescent modes become more prominent in the
vicinity of the source, their influence is reduced significantly by
using the average value of these two microphones for the
analysis. This technique helps to reduce the errors for frequencies
approaching the cut-on frequency of the first higher order mode and
thus, extends the frequency range for accurate results.

At the end of each section a loudspeaker is mounted at the
circumference of the duct ($A$ and $B$ in
Fig.~\ref{fig:test-rig}). They deliver the test signal for the damping
measurements. The signal used here is a multi-tone sine signal. All
tonal components of the signal are in the plane wave range. The signal
has been calibrated in a way that the amplitude of each tonal
component inside the duct is about 102\,dB.

The microphones used in these measurements are 1/4" G.R.A.S. type 40BP
condenser microphones. Their signals are recorded with a 16 track OROS
OR36 data acquisition system with a sampling frequency of
8192\,Hz. The source signals for the loudspeakers are recorded on the
remaining tracks. The test signal is produced by an Agilent 33220A
function generator. The signals are fed through a Dynacord L300
amplifier before they power the Monacor KU-516 speakers.

For each configuration two different sound fields are excited
consecutively in two separate measurements (index a and b). Speaker A
is used in the first measurement and in the second measurement the
same signal is fed into speaker B. Then, the data of section 1 and
section 2 (index 1 and 2) are analyzed separately. This results in
four equations for the complex sound pressure amplitudes for each
section and measurement for $j=a,b$:
\begin{subequations}
  \begin{align}
    \hat{p}_{1j}^{}(z)&=\hat{p}_{1j}^{+}e^{\imath
                        \omega z/c}+\hat{p}_{1j}^{-}e^{-\imath
                        \omega z/c}\label{eq:p1a}\\ 
    \hat{p}_{2j}^{}(z)&=\hat{p}_{2j}^{+}e^{\imath
                        \omega(z-L)/c}+\hat{p}_{2j}^{-}e^{-\imath
                        \omega(z-L)/c}\label{eq:p2b} 
  \end{align}
\end{subequations}
$\hat{p}^{+}$ and $\hat{p}^{-}$ are the complex amplitudes of the
downstream and upstream traveling waves.

The recorded microphone signals are transformed into the frequency
domain using the method presented by Chung~\cite{chung1977}. This
method rejects uncorrelated noise, \eg turbulent flow noise, from the
coherent sound pressure signals. Therefore, the sound pressure
spectrum of one microphone is determined by calculating the
cross-spectral densities between three signals, where one signal
serves as a phase reference. In our case the phase reference signal is
the source signal of the active loudspeaker. As a result we obtain a
phase-correlated complex sound pressure spectrum for each microphone
signal.

\begin{figure}[!bt]
  \centering
  \includegraphics[width=0.45\linewidth]{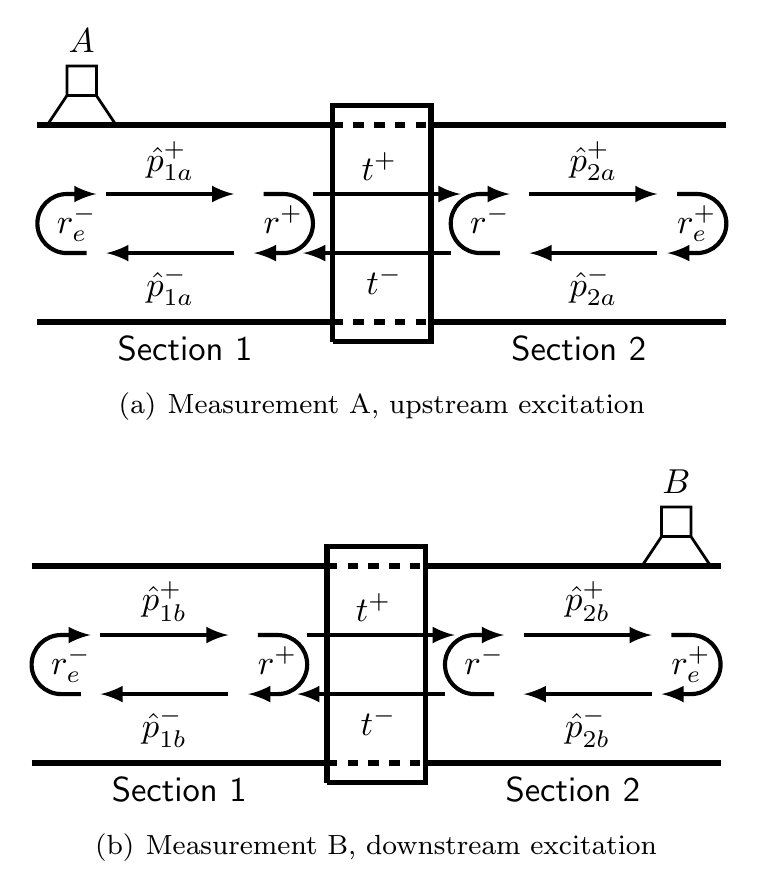}
  \caption{Illustration of the sound filed in the duct for
    measurements $A$ and $B$ by means of the sound pressure amplitudes
    $\hat{p}$, the reflection coefficient $r$, the transmission
    coefficient $t$, and the end reflection $r_e$}
  \label{fig:sound-field}
\end{figure}

According to Eqs.~\eqref{eq:p1a}-\eqref{eq:p2b} the measured acoustic signal is a superposition of two plane waves traveling in opposite direction. In order to determine the downstream and upstream propagating portions of the wave in each section, a mathematical model is fitted to the acoustic microphone data. This model considers viscous and thermal conductivity losses at the duct wall. They are included in the wave number with the following attenuation factor $\alpha$ as proposed by Kirchhoff~\cite{kirchhoff1868}:
\begin{equation}
\alpha=\frac{1}{c \Rd}\sqrt{\frac{\nu \omega}{2}}\left(1+\frac{\gamma-1}{\sqrt{Pr}}\right)
\end{equation}
with the duct radius $r$, the speed of sound $c$, the kinematic
viscosity $\nu$, the angular frequency $\omega$ (as in
  Eq.~\eqref{eq:Navier_Stokes}), the heat capacity ratio $\gamma$,
and the Prandtl number $Pr$.  As a result of this least-mean-square
fit, the four complex sound pressure amplitudes $\hat{p}_{1}^{+}$,
$\hat{p}_{1}^{-}$, $\hat{p}_{2}^{+}$ and $\hat{p}_{2}^{-}$ are
identified at position $z=0$ for both measurements. These sound
pressure amplitudes are related to each other via the reflection and
transmission coefficients of the test object. This is illustrated in
Fig.~\ref{fig:sound-field} for the two different measurements $A$ and
$B$. In order to calculate the reflection and transmission
coefficients $r^+$, $r^-$, $t^+$, and $t^-$ from the sound pressure
amplitudes the following four relations can be derived for $j=a,b$:
\begin{subequations}
  \begin{align}
    \hat{p}_{1j}^{-}=r^{+}\hat{p}_{1j}^{+}+t^{-}\hat{p}_{2j}^{-}\\
    \hat{p}_{2j}^{+}=r^{-}\hat{p}_{2j}^{-}+t^{+}\hat{p}_{1j}^{+}
  \end{align}
\end{subequations}

The equations from both measurements are combined and solved for the reflection 
\begin{equation}
r^{+}=\frac{\hat{p}_{1a}^{-}\hat{p}_{2b}^{-}-\hat{p}_{1b}^{-}\hat{p}_{2a}^{-}}{\hat{p}_{1a}^{+}\hat{p}_{2b}^{-}-\hat{p}_{1b}^{+}\hat{p}_{2a}^{-}}\quad\quad
r^{-}=\frac{\hat{p}_{2b}^{+}\hat{p}_{1a}^{+}-\hat{p}_{2a}^{+}\hat{p}_{1b}^{+}}{\hat{p}_{1a}^{+}\hat{p}_{2b}^{-}-\hat{p}_{1b}^{+}\hat{p}_{2a}^{-}}
\label{eq:ReflectionCoefficients}
\end{equation}
and transmission coefficients
\begin{equation}
t^{+}=\frac{\hat{p}_{2a}^{+}\hat{p}_{2b}^{-}-\hat{p}_{2b}^{+}\hat{p}_{2a}^{-}}{\hat{p}_{1a}^{+}\hat{p}_{2b}^{-}-\hat{p}_{1b}^{+}\hat{p}_{2a}^{-}}\quad\quad
t^{-}=\frac{\hat{p}_{1a}^{+}\hat{p}_{1b}^{-}-\hat{p}_{1b}^{+}\hat{p}_{1a}^{-}}{\hat{p}_{1a}^{+}\hat{p}_{2b}^{-}-\hat{p}_{1b}^{+}\hat{p}_{2a}^{-}}
\label{eq:TransmissionCoefficients}
\end{equation}
in downstream and upstream direction, respectively. The advantage of
combining the two measurements is that the resulting coefficients are
independent from the reflection of sound at the duct
terminations. These end-reflections are contained in the sound
pressure amplitudes, but do not need to be calculated
explicitly. Moreover in the case of a uniform and stagnant flow
  these coefficients do not depend on the direction we consider, \ie
  $r^-=r^+$ and $t^-=t^+$.

The dissipation of acoustic energy is expressed by the dissipation
  coefficient. The dissipation coefficient $\bDelta$ can be calculated
  directly from the reflection coefficient $R$ and the transmission
  coefficient $T$ via an energy balance
\begin{equation}
  \label{eq:EnergyConservation}
  R^{\pm}+T^{\pm}+ \bDelta^{\pm}=1.
\end{equation}
To compute these coefficients, the integration of the acoustic
  energy flux in a uniform and stagnant flow yields a relation between
  the acoustic pressure $p$ and acoustic power $P$ quantities(see
  Blokhintsev~\cite{blokhintsev1946} and Morfey~\cite{morfey1971})~:
\begin{equation}
  \label{eq:Power}
  P^{\pm}=\frac{\pi \Rd^2}{2\rho_0 c} \left|\hat{p}^{\pm}\right|^{2}
\end{equation}
Then, the energy coefficients can be given relative to the pressure
coefficients as: 
\begin{subequations}
  \label{eq:EnergyReflection}
  \begin{align}
    R^{+}
    &=\tfrac{P_{1}^{-}}{P_{1}^{+}}
      = \left|r^{+}\right|^{2}
      \label{eq:EnergyReflectionPlus}\\
    R^{-}
    &=\tfrac{P_{2}^{+}}{P_{2}^{-}}
      = \left|r^{-}\right|^{2}
      \label{eq:EnergyReflectionMinus}\\
    T^{+}
    &=\tfrac{P_{2}^{+}}{P_{1}^{+}}
      = \left|t^{+}\right|^{2}
      \label{eq:EnergyTransmissionPlus}\\
    T^{-}
    &=\tfrac{P_{1}^{-}}{P_{2}^{-}}
      = \left|t^{-}\right|^{2}
      \label{eq:EnergyTransmissionMinus}
  \end{align}
\end{subequations}
where the indices $1$ and $2$ refer to section 1 and section 2 of the
duct as illustrated in Fig.~\ref{fig:sound-field}. 
With the energy balance~\eqref{eq:EnergyConservation} follows the
  definition of the energy dissipation coefficient
\begin{equation}
  \bDelta = 1-\left(
  \left|r^{\pm}\right|^{2}+\left|t^{\pm}\right|^{2}\right)
\label{eq:dissipation}
\end{equation}
This is an integral value of the acoustic energy that is absorbed
while a sound wave is passing the damping module. The dissipation
coefficient is used to evaluate the damping performance of the test
object.


\subsubsection{\sffamily Numerical simulation of dissipation losses}

\begin{figure}[!bt]
  \centering
  \includegraphics[width=0.95\linewidth]{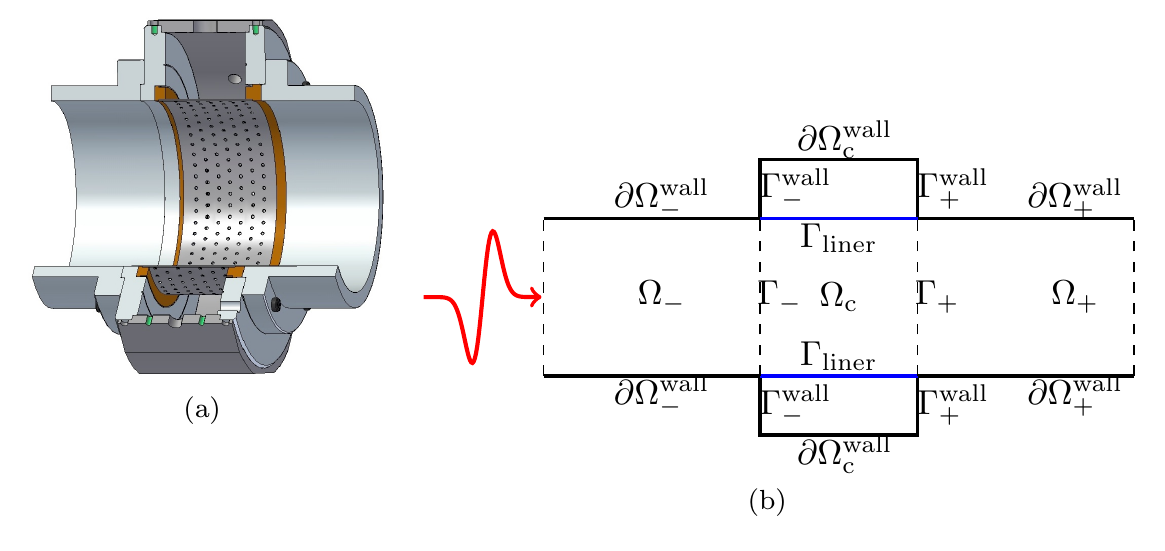}
  \caption{Split of the domain $\Omega$ into the two semi-infinite
    waveguides $\Omega_\pm$ and the multi-perforated liner section
    $\Omega_{\textrm{c}}$.  Impedance transmission conditions on the
    interface $\Gamma_\liner$ approximate the behaviour of
    the many perforations.}
  \label{fig:split}
\end{figure}

This setup is also simulated numerically using the equivalent
problem~\eqref{eq:Helmholtz:pressure}-\eqref{eq:impedance_conditions:pressure_only}
for the pressure with a source term corresponding to an incoming field
$p_\inc(r,\theta,z) = \exp(\imath \omega z / c)$ from the left. The
scattered field is computed numerically using the mode matching
procedure with $N=5$ modes~\cite{Semin.Thoens-Zueva.Schmidt:2017}: we
seek for the scattered field $p_0$ under the form (see
Fig.~\ref{fig:split}(b))
\begin{subequations}
  \label{eq:mode_decomp}
  \begin{align}
    p_0(r, \theta, z) & = p_\inc(r,\theta,z) + \sum_{j=0}^{N-1}
                        \alpha_j^- \,\psi_j(r)\, \exp(- \imath\, \beta_j\, z), \quad
                        z<0, \\
    p_0(r, \theta, z) & = \sum_{j=0}^{N-1}
                        \alpha_j^+ \,\psi_j(r)\, \exp(\imath\, \beta_j\, z), \quad
                        z>L,
  \end{align}
  inside the waveguide part, and under the form
  \begin{equation}
    \label{eq:mode_decomp_chamber}
    p_0(r, \theta, z) = \sum_{j=0}^{2N-1}
    \psi'_j(r)\big(
    {\alpha'_j}^+ \exp(\imath\, \beta_j'\, z)  + 
    {\alpha'_j}^-  \exp(\imath\, \beta_j'\, (L-z))\big), \quad 0 < z <
    L,    
  \end{equation}
  inside the duct part. The pairs $(\beta_j,\psi_j)$ and
  $(\beta_j',\psi_j')$ are solution of a ``2D'' transverse eigenvalue
  problem in the wave-guide and liner parts, using the fact that the source term $p_\inc$ and the
  geometry are independent of the angle~$\theta$. From the mode matching and
  assuming that there is only one propagative mode inside the
  waveguide, \ie $\beta_j \in \imath\IR$ for $j \not=0$, the energy
  dissipation coefficient is computed as
\end{subequations}
\begin{equation}
  \label{eq:dissipation_numerical}
  \bDelta := 1 - \left( |\alpha_0^+|^2 + |\alpha_0^-|^2 \right),
\end{equation}
and corresponds to the energy dissipation coefficient $D^\pm$
(Eq.~\eqref{eq:dissipation}) if both grazing and bias flows are absent.

\subsection{\sffamily Numerical results and comparison with experimental data}
\label{sec:numerical-results}

\begin{figure}[!bt]
  \centering
  \includegraphics[width=0.95\linewidth]{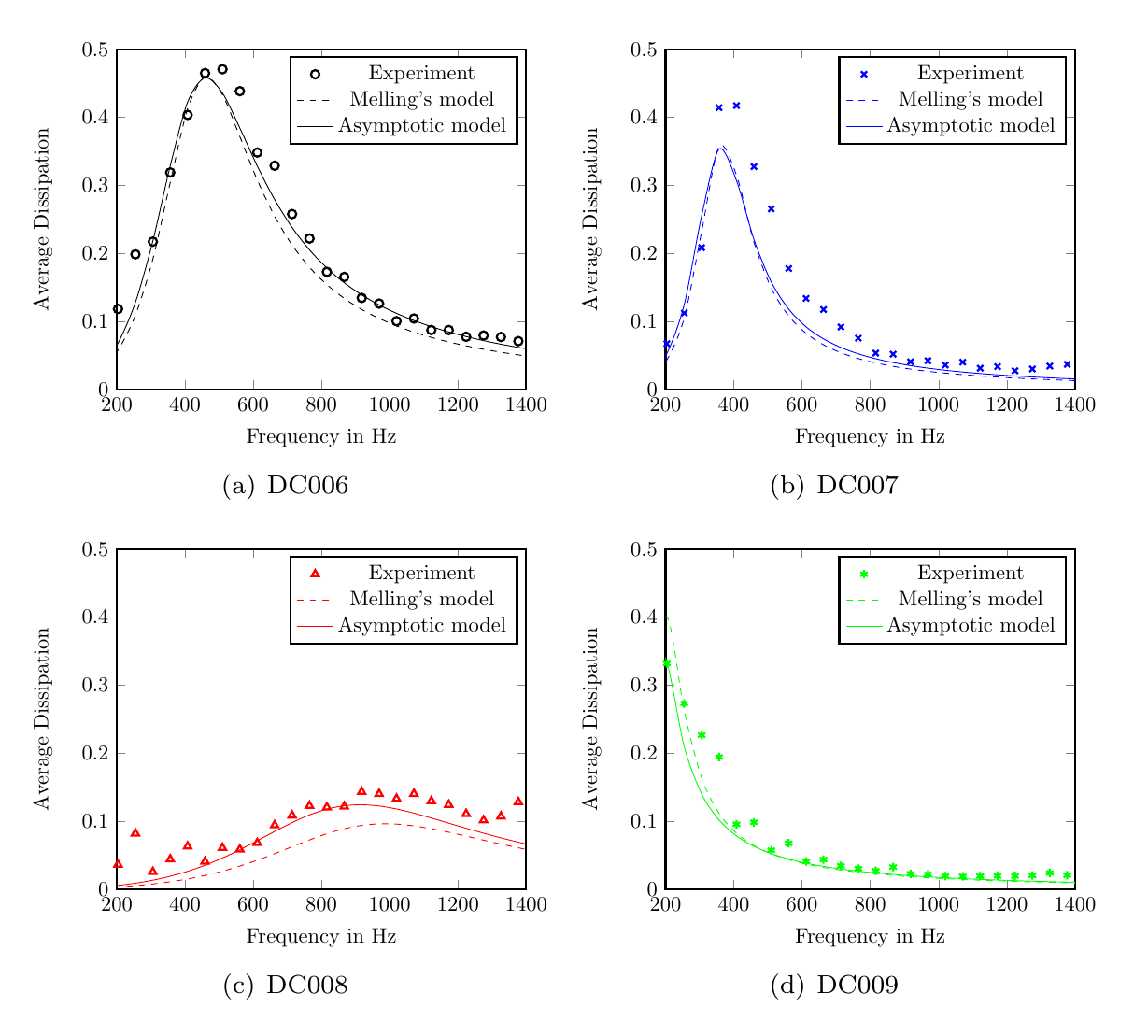}
  \caption[Compare dissipation]{%
    Average dissipation from experiments and numerical modelling plotted over the frequency comparing models for DC006, DC007, DC008, DC009.}
  \label{fig:CR}
\end{figure}

Figure~\ref{fig:CR} shows the average dissipation of the different liner configurations (see table~\ref{tab:liner_configurations}) in the DUCT-C setup (see figure~\ref{fig:sound-field}) as a function of the frequency. The average dissipation represents a mean value of the dissipation results for the upstream and downstream acoustic incidence (see section~\ref{sec:experimental-setup}). In a symmetric setup and without grazing flow this is, of course, equal to the dissipation from either side of excitation. The graphs compare the experimental values (symbols), the former theoretical model from Melling~\cite{melling1973acoustic} (dashed lines) and the here introduced asymptotic model (solid lines). 
In result, the asymptotic model indicates a better comparison to the experimental values especially for the configurations DC006 (figure~\ref{fig:CR} (a)) and DC008 (figure~\ref{fig:CR} (c)) where the Melling model slightly underestimates the dissipation in the frequency range above approximately 400~Hz. For the configuration DC007 with a porosity of 1.0~\% and a hole diameter of 2.5~mm both models (Melling and asymptotic) underestimate the maximum dissipation of approximately 0.4 around 400~Hz revealed in the experimental studies. 

\section{\sffamily Conclusions}

It has been shown that impedance conditions with one numerically computed parameter -- the effective Rayleigh conductivity -- 
can predict well the dissipation losses of acoustic liners. %
The effective Rayleigh conductivity can be obtained by solving
numerically an instationary Stokes problem in frequency domain
of one hole with a scaled viscosity in an characteristic infinite domain with prescribed pressure at infinity. %
For the computation the infinite domain is truncated, where we propose approximative boundary conditions on the artificial boundaries 
and an extrapolation procedure to save computation time. %
We decoupled in a systematic way the effects at different scales and derived impedance conditions for the macroscopic pressure or velocity 
 based on a proper matching of pressure and velocity at the different scales.
In difference to a direct numerical solution for the acoustic liner the overall computation effort is separated into a precomputation of the effective Rayleigh conductivity
and a computation of the Helmholtz equation for the pressure with impedance conditions, where no holes have to be resolved anymore by a finite element mesh.
The comparison with measurements in the duct acoustic test rig with a circular cross-section at DLR Berlin show that the dissipation losses based
on the impedance conditions with effective Rayleigh conductivity are well predicted.
The derivation of the impedance conditions do not depend on the cylindrical shape of the liner and can be used for others shapes like rectangular profiles.
The procedure for the computation of the effective Rayleigh conductivity can not only be extented to include thermic effects that are currently only heuristically incorporated,
but also nonlinear effects inside the hole that lead to an interaction of frequencies.

\section*{\sffamily Acknowledgements}

The authors would like to thank Claus Lahiri (Rolls-Royce) for
fruitful discussions.

The research was supported by Einstein Center for Mathematics Berlin
via the research center MATHEON, Mathematics for Key Technologies, in
Berlin as well as the Brandenburgische Technische Universit{\"a}t
Cottbus-Senftenberg through the Early Career Fellowship of the second
author.

The research was partly conducted during the stay of the first and
second author at the TU Berlin and the first author at BTU
Cottbus-Senftenberg.



\section*{\sffamily References}

\begingroup
\renewcommand{\section}[2]{}%

\endgroup

\end{document}